# Power-Efficient Image Storage: Leveraging Super Resolution Generative Adversarial Network for Sustainable Compression and Reduced Carbon Footprint

Satyam Singh
School of Electronics Engineering
Vellore Institute of Technology
Chennai, India
satyamsingh20027@gmail.com

Dr. Ashok Mondal
School of Electronics Engineering
Vellore Institute of Technology
Chennai, India
0000-0002-8410-9198

*Abstract*—In recent years, large-scale adoption of cloud storage solutions has revolutionized the way we think about digital data storage. However, the exponential increase in data volume, especially images, has raised environmental concerns regarding power and resource consumption, as well as the rising digital carbon footprint emissions. The aim of this research is to propose a methodology for cloud-based image storage by integrating image compression technology with Super-Resolution Generative Adversarial Networks (SRGAN). Rather than storing images in their original format directly on the cloud, our approach involves initially reducing the image size through compression and downsizing techniques before storage. Upon request, these compressed images will be retrieved and processed by SRGAN to generate images. The efficacy of the proposed method is evaluated in terms of PSNR and SSIM metrics. Additionally, a mathematical analysis is given to calculate power consumption and carbon footprint assesment. The proposed data compression technique provides a significant solution to achieve a reasonable trade off between environmental sustainability and industrial efficiency.

## I. Introduction

The rapid evolution of information and communication technologies (ICTs) in the digital domain has led to the widespread adoption of cloud storage services, resulting in a staggering increase in data generation regardless of its necessity [1]. Data centers, essential for maintaining this digital infrastructure, consume significant resources to operate efficiently. However, the exponential growth in digital content often stems from indiscriminate data storage practices, contributing to the substantial energy consumption of data centers, which presently accounts for approximately 1% of global electricity usage [2]. Moreover, the environmental repercussions of these data centers and cloud infrastructures are becoming increasingly concerning. The concept of the digital carbon footprint, representing the greenhouse gas emissions associated with devices, platforms, and digital technology resources [3], sheds light on how inefficient storage systems exacerbate global warming. Images, constituting a substantial portion of internet data, pose particular challenges due to their large file sizes, quantities and the high energy demands of the data centers housing them. In the UK alone, unwanted images contribute to 355,000 tonnes of $CO_2$ emissions annually [4], underscoring the environmental impact of unoptimized data storage practices. Research indicates that the indiscriminate accumulation of photos by individuals in Britain results in a larger carbon footprint compared to emissions from round-the-world flights [4]. Depending on the architecture of the cloud storage service used, for each kilowatt hour (KWH) of consumption, there's a carbon emission of approximately 500 grams(g) [5]. There have been numerous studies analyzing the carbon footprint generated due to the storage, usage, and transmission of data [6]. However, given the progressive nature of the internet sector, there is a continuous need to update these estimates [1]. As far as the volume of images is concerned, studies suggest that as of 2022, there are at least 136 billion indexable images on Google Images, indicative of the vast volume of digital content on the internet [7]. However, the actual total may be considerably higher, highlighting the magnitude of digital content generation and its associated environmental consequences [7].To tackle these challenges effectively, innovative solutions are required to mitigate the environmental footprint of digital image management, while simultaneously ensuring the functionality and accessibility of digital content. Our research aims to explore innovative approaches by integrating image compression algorithms for efficient storage and compression of images, coupled with Super- Resolution Generative Adversarial Network(SRGAN) based super-resolution techniques [8].This dual approach seeks to enhance storage efficiency by storing images in lesser storage size, thereby promoting environmental sustainability. SRGAN ensures that images can be restored to their original resolutions as needed, ensuring on-demand availability.

## II. DATASET

In our research, we employed the DIV2K dataset [9], which comprises a total of 1000 images. These images are of 2K resolution, ensuring they possess at least 2000 pixels on one of their axes. The dataset is split into three subsets: 800 images for training, 100 for validation, and another 100 for testing. Our focus lies in utilising high-resolution training images as well as four times(4x) downscaled images for training the model, aligning with our objective of assessing the performance of an SRGAN model tailored for 4x super-resolution enhancement. This approach ensures consistency in our evaluation process by utilizing downscaled images that match the enhancement factor provided by the SRGAN model. Moreover, as a part of our evaluation process, we utilized the Set5 dataset, renowned within the field for its compilation of five high-quality RGB images. This dataset is prominently featured in various studies and serves as a standard benchmark for assessing performance in image super-resolution tasks.

## III. THEORETICAL BACKGROUND OF SRGAN

SRGAN is a deep learning method that uses convolutional neural networks(CNNs) along with Generative Adversarial Networks(GANs). The SRGAN aims to convert low-resolution images into realistic high-resolution ones. GANs are based on the concpet of generator and discriminator. The discriminator aims to differentiate between the orginal ground truth images and the images created by the generator. Through this adversarial training process [10], the generator iteratively improves its ability to create exceptionally realistic high-resolution images. It achieves this by leveraging feedback from the discriminator and adjusting its approach through backpropagation to minimize loss.

### A. Generator Architecture

The generator architecture of the SRGAN model starts with a low-resolution input image, which undergoes convolution with 64 filters of size 9x9. The output is then passed through the parametric rectified linear unit (ReLU) function. This is followed by a series of 16 residual blocks, each containing convolutional layers with 64 filters of size 3x3, parametric ReLU, and batch normalization. The outputs of these blocks are combined with the original input through elementwise sum operations, facilitating feature extraction while preserving information. After passing through the residual blocks, the output undergoes another convolutional layer and batch normalization before being combined with the initial ReLU output through an elementwise sum. Subsequently, the upsampling process begins, utilizing pixel shuffling to gradually increase image resolution. This typically involves two upscaling blocks. Finally, the architecture concludes with a convolutional layer which is resposible for generating the high-resolution image as the final output. The generator architecture is shown in Fig. 1.

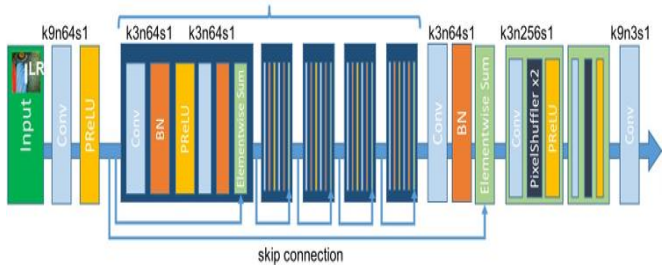

Fig. 1. SRGAN generator. The architecture details such as the number of feature maps (n),kernel size (k), and stride (s) are specified for each convolutional layer.

### B. Discriminator Architecture

The discriminator network of the SRGAN model operates as a Convolutional Neural Network (CNN) specialized in image classification. It distinguishes between authentic high-resolution images and those generated by the generator. Initially, input images, regardless of their origin, undergo convolutional layers to extract fundamental features. These features are then subjected to the leaky ReLU activation function for further processing. Subsequently, the features pass through multiple discriminator blocks, each comprising of batch Normalization, convolutional layer and leaky ReLU activation. These blocks enable the network to discern intricate features from the input images. Finally, the processed features are forwarded through Dense layers, followed by additional Leaky ReLU activation and Dense layers to produce the output. Given its classification objective, the discriminator network is trained to accurately classify images based on the extracted features, thereby effectively distinguishing between ground truth images and the images created by generator. The discriminator network is briefly described in Fig. 2.

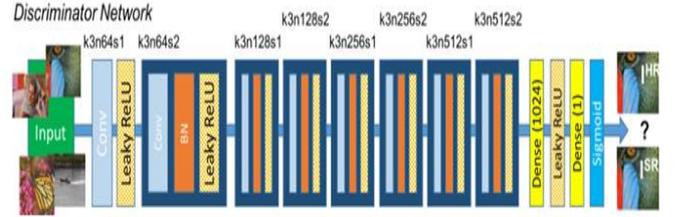

Fig. 2. SRGAN discriminator. The architecture details such as the number of feature maps (n),kernel size (k), and stride (s) are specified for each convolutional layer.

### C. Perceptual Loss

What makes SRGAN unique is the loss function that is used in the process. Before SRGAN, traditional loss functions relied heavily on the mean square error (MSE) between the generated and real images. While effective in ensuring a high signal-to-noise ratio, this method often led to a loss of significant high-frequency information. As a result, the generated images, although of high fidelity, lacked the intricate details crucial for true realism. SRGAN introduced a novel approach by utilizing a combination of loss functions. Firstly, it incorporated the Visual Geometry Group(VGG) or the content loss, computed by comparing feature representations extracted from multiple layers of a pre-trained VGG network. This loss, computed as the Euclidean distance between feature maps, helped preserve important details during training. Additionally, SRGAN employed adversarial loss, leveraging the discriminator's predictions. Here, the generator aimed to minimize this loss while the discriminator sought to maximize it. By integrating adversarial loss, SRGANs empowered the generator to produce super-resolved images closely resembling genuine high-resolution images in appearance. This combination of losses facilitated the generation of images with enhanced realism and detail, achieving a significant advancement in super-resolution image generation. The perceptual loss($L_{SR}$) function in SRGAN is given in (1) where content loss is given by $L_X^{SR}$ and adversarial loss is given by $L_{gen}^{SR}$.

$$L_{SR} = L_X^{SR} + 10^{-3} L_{Gen}^{SR} \qquad (1)$$

## IV. METHODOLOGY

The study is focused on the development of a novel technique to minimize the storage footprint of image data. To achieve our target, various process involved in image compression are used. The deflate compression algorithm, which combines LZ77 and Huffman coding [11], is applied using the Sharp library provided by the node package manager(npm) [12] for image compression. Additionally, we utilize adaptive filtering methods provided by the library to dynamically adjust the compression algorithm based on image characteristics. To further enhance compression rates, we implement Floyd-Steinberg dithering, a technique that reduces the color palette of an image while preserving perceived details [13]. Through this restricted color palette, the image size is reduced. The dithering algorithm employed in this study aims to minimize quantization errors by redistributing them to neighboring pixels that have not yet been processed. Although this approach results in some loss of information due to color palette restriction, the perceptual quality of the image remains high [13]. Following compression, we downscale the images using the Sharp library and Lanczos3 interpolation method, achieving a 4x downscaling to obtain the desired storage format. In the subsequent stage, our focus shifts to leveraging the SRGAN model to enhance the compressed and downscaled images obtained from the previous phase. The SRGAN model is trained on the DIV2K dataset, comprising 800 pairs of high-resolution and 4x downscaled low-resolution images. Training is conducted using the NVIDIA T4 GPU within the TensorFlow environment, facilitated by the TensorLayer library [14] [15]. To preserve high-level features during training, the perceptual loss is calculated using a pre-trained VGG19 network and then integrated into the model architecture. Following training, the processed, dowscaled images from inital stage are feeded into the SRGAN model, yielding super-resoluted outputs ready for evaluation.

## V. PROPOSED ARCHITECTURE

The depicted architecture, as illustrated in Figure 5, initiates with the acquisition of image data. Subsequently, these images undergo compression and Floyd-Steinberg dithering processes. After this, the images are 4x downscaled using lancoz3 interpolation. The resultant processed images are then ready for storage in systems, occupying much reduced storage space compared to their original sizes. Upon demand, images are retrieved from storage and forwarded to the SRGAN model for super-resolution. Ultimately, the end user receives the original resolution image for utilization.

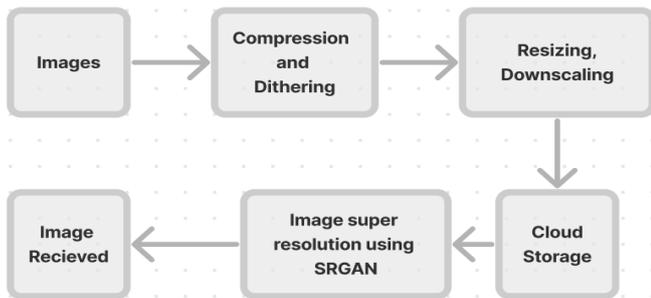

Fig. 3. Block diagram of proposed architecture

## VI. RESULTS

In our evaluation, we used validation images sourced from DIV2K dataset. Apart from this, we're also using SET5 dataset. We're using the same compression and downscaling method as proposed in Fig3 before giving the images to SRGAN for evaluation. While conventional metrics like PSNR and SSIM are considered, our research extends beyond these to include an emphasis on compression ratios. PSNR stands for peak signal-to-noise ratio and represents the ratio of maximum possible power output to the power of noise which impacts the quality of image. SSIM stands for Structural Similarity index and is an indicator of image quality. Furthermore, we estimate the potential environmental benefits, particularly in terms of reduced digital carbon footprint and power usage. This approach allows us to underscore the method's efficacy through a sustainability lens. While maintaining uniformity in the compression and downscaling procedures across all images, we vary the levels of Floyd-Steinberg dithering to observe their impact. Specifically, within the Sharp library, the dithering level is adjustable on a scale from 0 to 1, with 1 representing the highest dithering intensity. The initial file size of Set5 dataset is 0.81 Megabytes(MBs) while that of DIV2K validation images dataset is 428 MB. Table I draws comparison with PSNR, SSIM and compression metrics of DIV2K and Set5 dataset with different dithering scales. The dataset file sizes are given in megabytes in table and represent the downscaled images dataset sizes that we propose to store in cloud storage. It can be seen that for Set5, PSNR is approximately 27.22 db for both dither scales 0.5 and 1. For DIV2K, the PSNR values are 26.86 db and 26.87 db for dither scales 1 and 0.5

TABLE I. DITHERING SCALES COMPARISON ON SET5 AND DIV2K

| Dataset | Dither | PSNR | SSIM | Stored size | Compression percentage |
|---|---|---|---|---|---|
| DIV2K | 1 | 26.87 db | 0.77033 | 38.7 | 90.9579 |
| DIV2K | 0.5 | 26.86 db | 0.77019 | 38.5 | 91 |
| Set5 | 1 | 27.22 db | 0.82127 | 0.09140625 | 88.73 |
| Set5 | 0.5 | 27.22 db | 0.82126 | 0.0913 | 88.74 |

Respectively. The SSIM is approximately around 0.77 for DIV2K dataset for both dither scales and 0.82 for Set5 dataset dither scales 0.5 and 1. Compression percentage rates range between 88-91 percent for both the datasets. Fig. 4 illustrates a comparison between the original and generated SRGAN image from 0.5 dither level compression and downscaling. An analyis of the diference in original and generated SRGAN image used in Fig. 4 is depicted in Fig. 5. Furthermore, the power consumption of cloud storage per terabyte (TB) of data is estimated to range from 2.55 watts for distributed cloud storage to 11.55 watts for centralised cloud storage systems [12]. If we consider a data storage in TB units of data for a year on cloud storage, the energy consumption in kilowatt-hour(KWH) per year can be calculated for distributed ($E_{distributed}$) as well as centralised ($E_{centralised}$) storage systems as shown in (2) and (3), where **S** refers to the data stored on the cloud in TB units.

$$E_{distributed} = 2.55 \times 365 \times 24 \times S \qquad (2)$$

$$E_{centralised} = 11.55 \times 365 \times 24 \times S \qquad (3)$$

On using equations (2) and (3) on DIV2K images set with dither scale 1, we see that the original and final values of $E_{centralised}$ and $E_{distributed}$ corresponding to the energy usage of original(428MB) and final(38.7MB) file cloud storage sizes are as follows:-

- $E_{distributed}$ **initial** $\approx 9.118 \times 10^{-3}$ KWH
- $E_{distributed}$ **final** $\approx 0.824 \times 10^{-3}$ KWH

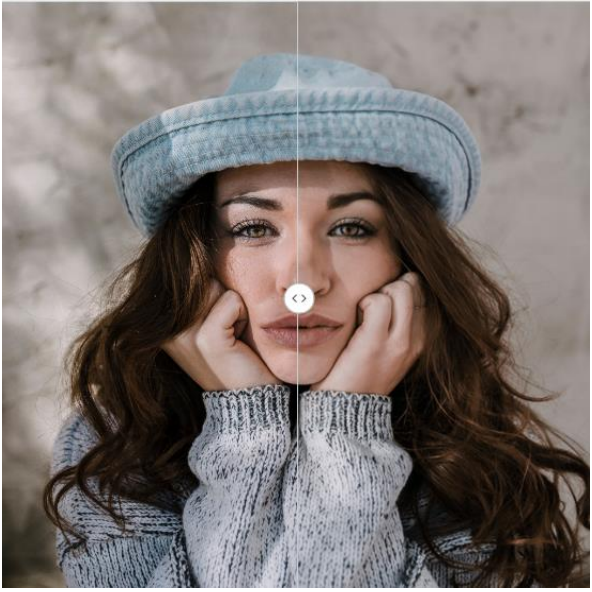

Fig. 4. DIV2K image 0855.PNG comparison: Original (Left-half) vs. SRGAN generated (Right-half, 0.5 Dithering)

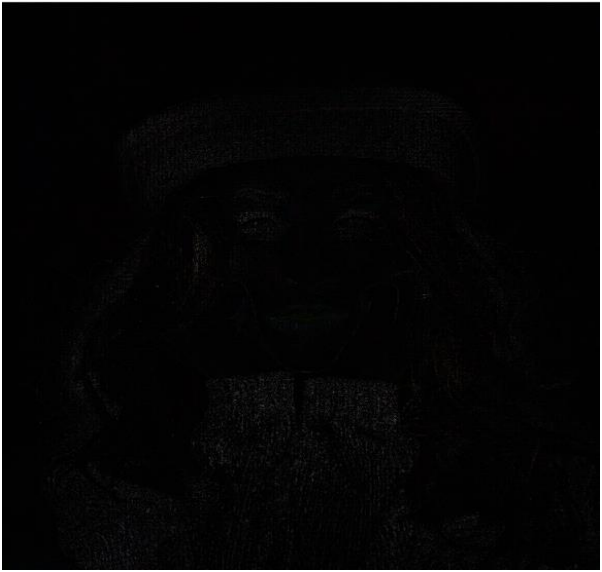

Fig. 5. DIV2K image 0855.PNG comparison: Overall comparsion between original image(left) and SRGAN generated (0.5 dithering) image

- $E_{centralised}$ initial $\approx 41.298 \times 10^{-3}$ KWH
- $E_{centralised}$ final $\approx 3.734 \times 10^{-3}$ KWH
- $E_{distributed}$ initial - $E_{distributed}$ final $\approx 8.294 \times 10^{-3}$ KWH
- $E_{centralised}$ initial - $E_{centralised}$ final $\approx 37.564 \times 10^{-3}$ KWH

The net energy savings observed annually on this dataset are approximately $8.294 \times 10^{-3}$ KWH for distributed storage and $37.564 \times 10^{-3}$ KWH for centralized storage. Using a conversion factor of 500 grams of carbon dioxide released per KWH of energy consumption [5], the annual reduction in carbon emissions using this storage strategy is estimated to be around 4.147 grams for distributed storage and 18.782 grams for centralized storage. Although these savings may appear modest given the small size of the DIV2K validation image dataset i.e 428MB, they become substantial when considering the operation of cloud storage systems at the terabyte or petabyte scale. Our approach has demonstrated compression percentages exceeding 90%, as depicted in Table 1. For instance, assuming a conservative estimate of 70% compression on image data with an original size of 10TB, the net yearly savings range from 156.366 KWH to 708.246 KWH depending on the storage system utilized. i.e centralised or distributed. Expanding this scenario, the corresponding reduction in carbon emissions is estimated to be between 78.183 kilograms and 354.123 kilogramsthrough storage system optimizations alone. Beyond environmental benefits and storage optimization, this approach renders storage cost-effective, presenting compelling advantages for industries and organizations.

## VII. CONCLUSION

Our research has demonstrated the significant potential of compression techniques combined with SRGAN in mitigating energy consumption and reducing carbon emissions in cloud storage systems. Our findings indicate that even modest reductions in data storage requirements can yield substantial energy savings and environmental benefits, particularly when considering the operation of cloud storage systems at scale. By implementing compression strategies that can achieve compression percentages even exceeding 90%, we have shown that annual energy savings and substantial corresponding reductions in carbon emissions can be realized depending on the storage system utilized. These results underscore the importance of considering environmental sustainability alongside storage optimization efforts in digital data management. Beyond the environmental benefits, our approach offers cost-effective storage solutions, providing compelling advantages for industries and organizations seeking to reduce their carbon footprint and operational costs. Our research highlights the scalability and practicality of our techniques in addressing the challenges posed by the growth of digital data.

## REFERENCES

[1] D. A. Kez, A. Foley, D. Laverty, D. D. F. Del Rio, and B. K. Sovacool, "Exploring the sustainability challenges facing digitalization and internet data centers,"Journal of Cleaner Production, vol. 371, p. 133633, Oct. 2022,

[2] Masanet, Eric & Shehabi, Arman & Lei, Nuoa & Smith, Sarah & Koomey, Jonathan,"Recalibrating global data center energy-use estimates",Science. 367. 984-986. 10.1126/science.aba3758,2020

[3] C. Heinze, "How to lower IT's digital carbon footprint," CIO, https://www.techtarget.com/searchcio/feature/Understand-the-digital-carbon-footprint-of-enterprise-IT(accessed Jan. 5,2024)

[4] "Dirty Data", IET, https://www.theiet.org/media/press-releases/press-releases-2021/press-releases-2021-october-december/26-october-2021-dirty-data/(accessed Feb. 10,2024)

[5] L. Posani, "The carbon footprint of distributed cloud storage," arXiv.org, https://arxiv.org/abs/1803.06973(accesed 15. Feb,2024)

[6] J. Aslan, K. Mayers, J. Koomey, and C. France, "Electricity intensity of internet data transmission: Untangling the estimates," Journal of Industrial Ecology, vol. 22, no. 4, pp. 785–798, Aug. 2017

[7] M. Broz, "How Many Photos Are There? (Statistics & Trends in 2024)",https://phototutorial.com/photos-statistics/(accesed 15 Feb,2024)

[8] C. Ledig, et al., "Photo-Realistic Single Image Super-Resolution Using a Generative Adversarial Network," 2017 IEEE Conference on Computer Vision and Pattern Recognition (CVPR), Honolulu, HI, USA,2017,pp.105-114.

[9] E. Agustsson and R. Timofte, "NTIRE 2017 Challenge on Single Image Super-Resolution: Dataset and Study,"2017 IEEE Conference on Computer Vision and Pattern Recognition Workshops (CVPRW), Honolulu, HI, USA, 2017, pp. 1122-1131, doi: 10.1109/CVPRW.2017.150

[10] Z. K. and A. Madry, "Chapter 4 Adversarial training, solving the outerminimization.",adversarial-ml-tutorial.org,https://adversarial-ml-tutorial.org/adversarial_training/ (accessed 5 Feb. 2024)

[11] "An explanation of the Deflate algorithm.", zlib.net ,https://www.zlib.net/feldspar.html (accessed Jan. 15,2024)

[12] "sharp high performance Node.js image processing.", https://sharp.pixelplumbing.com/ (accessed Jan. 25,2024)

[13] S. Sharma, J. J. Zou and G. Fang, "Detail and contrast enhancement for images using dithering based on complex wavelets", 2016 IEEE Region 10 Conference (TENCON), Singapore, 2016, pp. 1388-1391.

[14] Dong H, Supratak A, Mai L, Liu F, Oehmichen A, Yu S and Guo Y",TensorLayer: a versatile library for efficient deep learning development",ACM Int. Conf. on Multimedia,2017,pp. 1201–1204

[15] C. Lai, J. Han and H. Dong, "Tensorlayer 3.0: A Deep Learning Library Compatible With Multiple Backends,"2021 IEEE International Conference on Multimedia & Expo Workshops (ICMEW), Shenzhen, China, 2021, pp. 1-3